\documentclass[preprint,12pt]{elsarticle}
\usepackage{amssymb}
\usepackage{graphicx}

\newcommand{\nuc}[2]{\ensuremath{^{#1}}#2}

\newcommand{\AM}{$A\,$MeV}
\newcommand{\Zmoy}{\langle Z \rangle}

\journal{Physics Letters B}

\begin{document}

\title{
Phase transition dynamics for hot nuclei\\ 
INDRA Collaboration}

\author[ipno]{B.~Borderie}
\author[lpc]{N. Le Neindre}
\author[ipno]{M.~F.~Rivet\fnref{fn1}}
\author[csnsm]{P.~D\'esesquelles}
\author[suba]{E.~Bonnet}
\author[lpc]{R.~Bougault}
\author[ganil]{A.~Chbihi}
\author[ipno,nap]{D.~Dell'Aquila}
\author[ganil]{Q.~Fable}
\author[ganil]{J.D.~Frankland}
\author[ipno,cnam]{E.~Galichet}
\author[lpc]{D.~Gruyer}
\author[lyon]{D.~Guinet}
\author[nap]{M.~La~Commara}
\author[cat]{I.~Lombardo}
\author[lpc]{O.~Lopez}
\author[lpc,cher]{L.~Manduci}
\author[ipno]{P.~Napolitani}
\author[lpc,ifin]{M.~P\^arlog}
\author[nap]{E.~Rosato\fnref{fn1}}
\author[laval]{R.~Roy}
\author[ganil,laval]{P.~St-Onge}
\author[cat]{G.~Verde}
\author[lpc]{E.~Vient}
\author[nap]{M.~Vigilante}
\author[ganil]{J.P.~Wieleczko}
\fntext[fn1]{deceased}
\address[ipno]{Institut de Physique Nucl\'eaire, CNRS/IN2P3,
Univ. Paris-Sud, Universit\'e Paris-Saclay, Orsay, France}
\address[lpc]{LPC, CNRS/IN2P3, Ensicaen, Universit\'e de Caen, Caen, France}
\address[csnsm]{CSNSM, CNRS/IN2P3, Univ. Paris-Sud, Universit\'e
Paris-Saclay, Orsay, France}
\address[suba]{SUBATECH, IMT Atlantique, Universit\'e de
Nantes, CNRS-IN2P3, Nantes, France}
\address[ganil]{GANIL, (CEA/DRF-CNRS/IN2P3), Caen, France}
\address[nap]{Dipartimento di Fisica `E. Pancini' and Sezione INFN,
Universit\`a di  Napoli ``Federico~II'', Napoli, Italy}
\address[cnam]{Conservatoire National des Arts et M\'etiers, Paris, France}
\address[lyon]{Institut de Physique Nucl\'eaire, CNRS/IN2P3, Univ. Claude
Bernard Lyon 1, Universit\'e de Lyon, Villeurbanne France}
\address[cat]{INFN, Sezione di Catania, Catania, Italy}
\address[cher]{Ecole des Applications Militaires de l'Energie Atomique,
Cherbourg, France}
\address[ifin]{Hulubei National Institute for R \& D in Physics and
Nuclear Engineering (IFIN-HH), Bucharest-M\'agurele, Romania}
\address[laval]{Universit\'e Laval, Qu\'ebec, Canada}

\begin{abstract}

An abnormal production of events with almost equal-sized
fragments was theoretically proposed as a signature of spinodal
instabilities responsible for nuclear multifragmentation in the Fermi
energy domain. On the other hand finite size effects are predicted to
strongly reduce this abnormal production. High statistics quasifusion hot nuclei
produced in central collisions between $Xe$ and $Sn$ isotopes at 
32 and 45~\AM{} incident energies have been used to definitively
establish, through the experimental measurement of charge correlations,
the presence of spinodal instabilities. N/Z influence was also studied.

\end{abstract}



\maketitle

 During the last decades the thermodynamics of finite nuclear systems was widely studied with
 heavy-ion collisions at intermediate and relativistic energies and
 with hadron-nucleus collisions 
 at relativistic energies~\cite{WCI06,Bor08}. With such collisions,
 depending on impact parameter, a nuclear system can be
 heated, compressed and then diluted.
 These systems are expected to undergo a 
 liquid-gas type phase transition that manifests itself
 through nuclear multifragmentation~\cite{Bon85}. This theoretical expectation, 
 discussed for many years for nuclear matter~\cite{Sch82,Cur83,Jaq83,Mul95}
 is due to the similarity between the nuclear
 interaction and the Van der Waals forces acting in classical
 fluids~\cite{Jaq83,I46-Bor02}. 
 However, a nucleus (or a nuclear system) is a finite system which shows
 specific behaviours in the transition region.  
 Most of the predicted specific signals of phase transition are a direct consequence of
 the local convexity of the entropy which is expected for finite systems
 having a discontinuous transition in the thermodynamic
 limit~\cite{Wal94,Cho04,Gul03,WCI06}. By considering the microcanonical
 ensemble with energy as extensive
 variable, the convex intruder implies a backbending in the temperature
 (first derivative of entropy) at constant pressure and correlatively a 
 negative branch for the heat capacity (second derivative).
 Experimentally, these two converging signatures have been observed
 from very different analyses.
  Negative heat capacities with a microcanonical sampling, were observed for
 35~\AM{} Au+Au  semi-peripheral collisions~\cite{MDA00} and confirmed
 by the INDRA collaboration for 32-50~\AM{} Xe+Sn central
 collisions~\cite{I46-Bor02}. 
 For caloric curves, their shape depends on the
path followed by the system on the microcanonical equation of state
surface, and a backbending
(direct signature) can only be observed for a transition at constant
pressure~\cite{Cho00}. This was evidenced very recently for central
32-50~\AM{} Xe+Sn collisions, thanks to a simulation
based on experimental data~\cite{I66-Pia08} where
a quantal temperature was calculated from the
momentum fluctuations of protons present at freeze-out~\cite{Zhe11}.  
Pressure and volume-constrained caloric curves could be built and the expected
behaviours were observed:
a backbending for selected ranges of
pressure and a monotonous increase at constant average volume~\cite{I79-Bor13}. 
Another consequence of the entropy curvature anomaly manifests itself
when systems are treated in the canonical ensemble. 
In this case a direct phase transition
signature is the presence of a bimodal distribution of an order
parameter~\cite{Cho01} like 
the charge (size) of the largest fragment (Z$_{max}$) of multifragmentation
partitions. Bimodality was observed (with a canonical sampling) 
in 60-100~\AM{} Au+Au semi-peripheral
collisions, allowing moreover to estimate the latent heat for nuclei close to
gold around 8~MeV per nucleon and to set the appearance of the pure gas phase above
9-10 ~MeV per nucleon excitation energy~\cite{I72-Bon09}.

 The answer to a key point was still
 pending: it concerns the nature of the dynamics of
the transition, i.e. fragment formation. Two mechanisms have
been proposed. On one side, stochastic mean field approaches 
predict the fragmentation process to follow the spinodal fragmentation scenario
triggered by phase-space fluctuations amplified in an unstable medium and, on the
other side, in molecular dynamics models (QMD, AMD) many-body correlations
are sufficient to produce fragments at early times even in absence of 
unstable conditions~\cite{Cho04,Har98,I16-Neb99,Ono99,Ono02,I29-Fra01,Riz07}. 
Experimentally, there was an indication that multifragmentation
 may be induced by  spinodal instabilities but the confidence level of
 the fossil signature was not sufficient (3-4 $\sigma$), due to low statistics,
 to allow drawing any definitive conclusion~\cite{I31-Bor01,I40-Tab03,WCI-Bor06}.
 Such instabilities may occur when the system evolves through the
 mechanically unstable spinodal region of the phase diagram, located at densities
 $\rho\leq\rho_0$ and temperatures below the critical temperature. Such conditions are
 well explored in central nuclear collisions around Fermi energy~\cite{Bon14}.
 Moreover, if spinodal instabilities are at the origin of
 fragmentation, a reduction of 
 instabilities for N/Z asymmetric systems in relation with an increase of
 the instability growth time is theoretically predicted~\cite{Col02}.   
 In this letter we shall describe studies obtained with very high
 statistics ( a factor 10 to 15 higher as compared to previous experiments)
 aiming to give a final answer as far as spinodal
 fragmentation is concerned and, secondly, to search for the related isospin
 effects. 
 
 The experiment was performed at GANIL (Grand Acc\'el\'erateur
 National d'Ions Lourds) and two reactions were used 
 \nuc{124,136}{Xe}+\nuc{112,124}{Sn} at two bombarding energies, 32 and 45~\AM,
The beam, impinging on thin targets (530$\mu$g $cm^{-2}$), had an
intensity of about 3-5 $10^7$ ions per second to avoid event pile-up. 
Experimental data were collected with the 4$\pi$
multidetector INDRA which is described in detail in Refs.~\cite{I3-Pou95,I5-Pou96}.
Accurate particle and fragment identifications were achieved
and the energy of the detected products was measured
with an accuracy of 4\%. Further details can be found 
in Refs.~\cite{I14-Tab99,I33-Par02,I34-Par02}.

The data used in the analysis were obtained with an on-line
multiplicity trigger of 4 or more detected reaction products.
The number of such recorded events was between 60 and 80 million
events for each colliding system.
Quasi-complete events are selected by requiring that at least 80\% of the 
total charge of the system is measured. We then isolate compact
shape events (quasifusion) through the 
additional condition that the flow angle ($\theta_{flow}$) be larger than 
$60^{\circ}$. Let us recall that $\theta_{flow}$ characterizes the main
direction of matter emission in the center of mass of the reaction 
and is determined by the kinetic energy flow tensor calculated from
fragments (Z$\geqslant$5)~\cite{I28-Fra01}. 
Measured cross-sections corresponding to selected events are
$\approx$ 40~mb at 32~\AM{} and 25~mb at 45~\AM{}. They were derived
from the measured target thicknesses, the counting of ions collected in
the Faraday cup located at the end of the beam line and the
acquisition dead time. The charge of ions reaching the cup was
obtained using the formulas of Ref.~\cite{Sch01}. 
Total cross-sections for 
quasifusion events, taking into account detection efficiency and selection biases, 
are estimated to be $\sim$ 250~mb at 32~\AM{}, 
and $\sim$ 180~mb at 45~\AM{}.

In infinite nuclear matter the signature of spinodal instabilities is the formation
of equal-sized fragments due to density fluctuations which grow
exponentially with time. The most unstable modes correspond to
wavelengths lying around $\lambda \approx$ 10 fm and the associated
characteristic times are equal to around 30-50 fm/c,
depending on density ($\rho_0$/2 - $\rho_0$/8) and 
temperature (0-9 MeV)~\cite{Colo97,Nap17}. A direct consequence
of the dispersion relation is the production of ``primitive'' fragments
with size $\lambda$/2 $\approx$ 5 fm which correspond to $Z$ $\approx$
8 - 10 .
However, this simple picture is expected to be largely blurred by several
effects.
The beating of different modes occurs. Coalescence effects due to the
nuclear interaction between fragments before the complete disassembly
are also expected. For finite systems the situation is even more
complicated. The presence of a surface introduces an explicit breaking
of the translational symmetry with the important result that the
growth rates are nearly the same for different multipolarities,
which indicates that the unstable finite system breaks into different
channels depending on multipolarity $L$~\cite{Jac96}.
Equal-sized ``primitive'' fragments
are then expected to be produced with sizes in the range
$A_F$/2 - $A_F$/$L_{max}$; $A_F$ being the part of the system leading
to fragments during the spinodal fragmentation.
Moreover the finite system produced during the nucleus-nucleus
collision has to stay or live long enough in the spinodal region ($\sim$
3 characteristic time -100-150 fm/c - for symmetric matter) to allow
an important amplification of the initial density fluctuations.
And finally, we experimentally detect fragments after secondary decay,
which introduces a broadening of the fragment size distribution.
Taking into account the accumulation of all these effects on the final
extra production of equal-sized fragments, it is clear that any
signature that spinodal fragmentation is responsible for the phase
transition dynamics can only be what we have called a fossil signature.
A full simulation of the spinodal decomposition of quasifused sources
using Brownian One-Body dynamics calculations~\cite{I40-Tab03} already
testified to this fact, with less than 1\% of events with equal-sized
fragments. It is the reason why the signature is difficult to observe
experimentally.

Twenty years ago an intra-event correlation function called higher order charge
correlations~\cite{Mor96} was proposed to enlighten any extra production of
events with specific fragment partitions. The high sensitivity of the method
makes it particularly appropriate to look for small numbers of events as those
expected to have kept a memory of spinodal fragmentation properties.

\begin{figure}
\begin{center}
\includegraphics*[width=0.8\textwidth]
{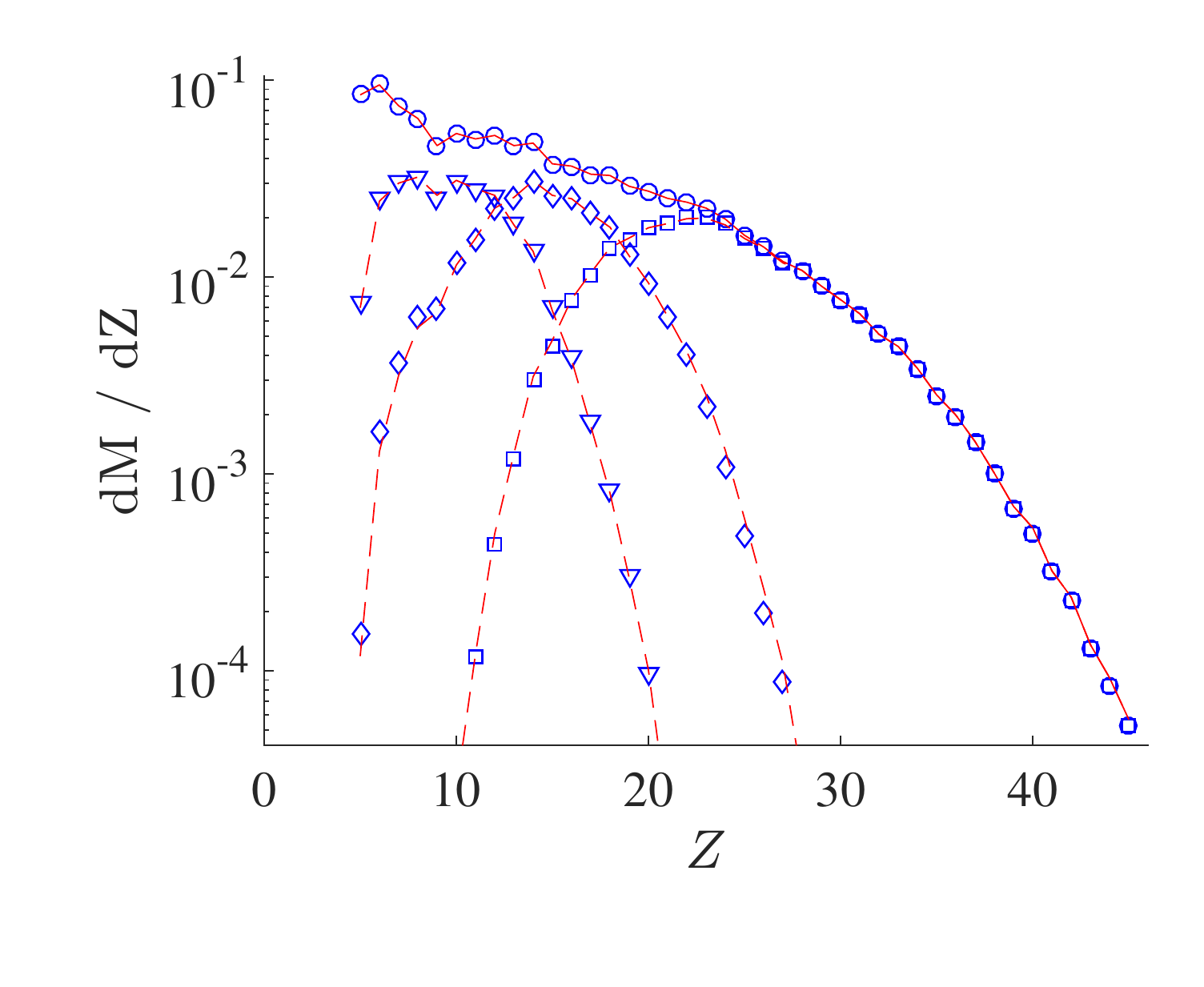}
\end{center}
\caption{(Color online) Experimental differential charge multiplicity
distribution (circles) for the single source formed in central 32~\AM{}
 \nuc{124}{Xe}+\nuc{112}{Sn} reaction with fragment multiplicity equal
to 4. Experimental differential distributions for the first (squares), second (diamond)
and third (triangles) heaviest fragments of partitions are presented too. The full
and dashed lines, to be compared to data, correspond to the results of the intrinsic
probability method for the fragment probabilities.}
\label{fig1}
\end{figure}

 All fragments of one event with fragment
multiplicity $M = \sum_Z n_Z$, where $n_Z$ is the number of fragments
with charge $Z$ in the partition,  are taken into account. By means of the
normalized first order:
\begin{equation}
	\Zmoy = \frac{1}{M} \sum_Z n_Z Z
\label{equ1}
\end{equation}
and second order:
\begin{equation}
        \sigma_Z^2 = \frac{1}{M} \sum_Z n_Z (Z - \Zmoy)^2
\label{equ2}
\end{equation}
moments of the fragment charge distribution in the event, 
one may define the correlation function (CF):
\begin{equation}
\left. 1+R(\sigma_Z, \Zmoy)=\frac{Y(\sigma_Z, \Zmoy)}{Y'(\sigma_Z, \Zmoy)} 
\right| _{M}
\label{equ3}
\end{equation}
Here, the numerator $Y(\sigma_Z, \Zmoy)$ is the yield of events with given
$\Zmoy$  and $\sigma_Z$ values. Because the measurement of the charge 
belonging to a given event is not subject to statistical fluctuations, we 
use here expression  (\ref{equ2}) rather than the ``nonbiased estimator'' 
of the variance,  $\frac{1}{M-1} \sum_Z n_Z (Z - \Zmoy)^2$,  as proposed in
~\cite{Mor96}.
The denominator $Y'(\sigma_Z, \Zmoy)$ represents the uncorrelated yield.
Its retained evaluation is based on the ``intrinsic probability'', $^{\rm intr}P_Z$,
of emission of a given charge. 
A unique set of intrinsic probabilities exists only if the chosen
event sample is extracted from single-source events. 
The denominator partition probabilities are built to take into account
in a combinatorial way 
the trivial correlations due to charge conservation.
Thus, this denominator replicates all features of the partitions of the numerator,
except the correlations due to other reasons than charge conservation.
The probability to observe 
a given partition (${\bf n}: (n_1,\ldots ,n_{Z_{\rm max}})$), at a given total
multiplicity, $m = \sum_Z n_Z$, is obtained by the multinomial 
formula. If the total charge is fixed ($Z_{\rm tot}=\sum_Z Z\,n_Z$), 
the partition probabilities are given by:
\begin{equation}
\label{Eq P(n)}
P({\bf n}|m) = \alpha \ m! \,
\prod_Z \frac{^{\rm intr}P_Z^{n_Z}} {n_Z!} \
\delta_{Z_{\rm tot},\sum_Z Z\,n_Z}\ ,
\end{equation}
\noindent
where $\alpha$ is the normalization constant  (so that $\sum_n
P({\bf n}|m$) = 1) and $\delta$ is the Kronecker symbol. 
All the details can be found in Refs.~\cite{I40-Tab03,Des02}.
The $^{\rm intr}P_Z$ values are obtained by means of a recursive
procedure of minimization which stops when the decreasing
between two steps is lower than 10$^{-12}$.
The minimization criterion is the normalized $\chi^2$
between the experimental partition probabilities and the calculated
partition probabilities for a considered set of data. 
The calculated $\chi^2$/$N_{dof}$ were always in the range 1.0-0.7: the lower
is the incident energy and the larger is the
multiplicity, the lower is the $\chi^2$ value. Fig.~\ref{fig1}
shows one example on how the experimental charge distributions are
faithfully described. The intrinsic probabilities,  
$^{\rm intr}P_Z$, have been calculated independently
for each incident energy, for each reaction and for the different
fragment multiplicities. Finally, to evaluate the sensitivity
of the intrinsic probability method for measuring CFs, we have
performed a Monte-Carlo simulation. The created events obey the
following conditions: as for data, events are selected by requiring at
least 80\% of the charge of a system with $Z$=104, charged reaction
products are produced according to the probability 
P(Z)~$\propto~\exp(-0.1Z)$ and fragments have Z$\geqq$5. $M$ was chosen 
equal to 4. As expected, CF values are close to one everywhere and, 
for low $\sigma_Z$ ($<$2), calculated CF
values are in the range 0.95-1.05; this last value determines the limit of
sensitivity.

\begin{figure}
\begin{center}
\includegraphics*[width=0.48\textwidth]
{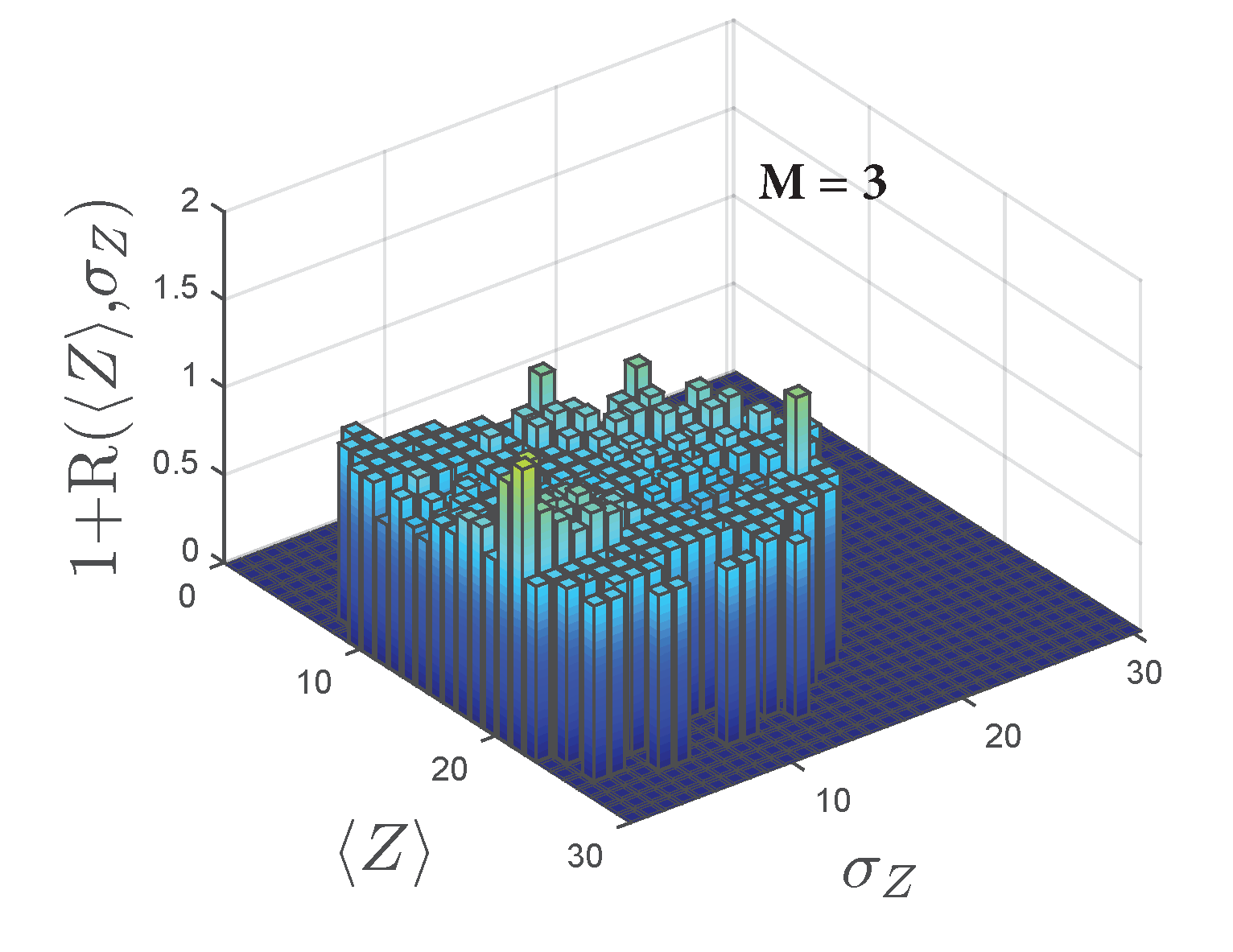}
\includegraphics*[width=0.48\textwidth]
{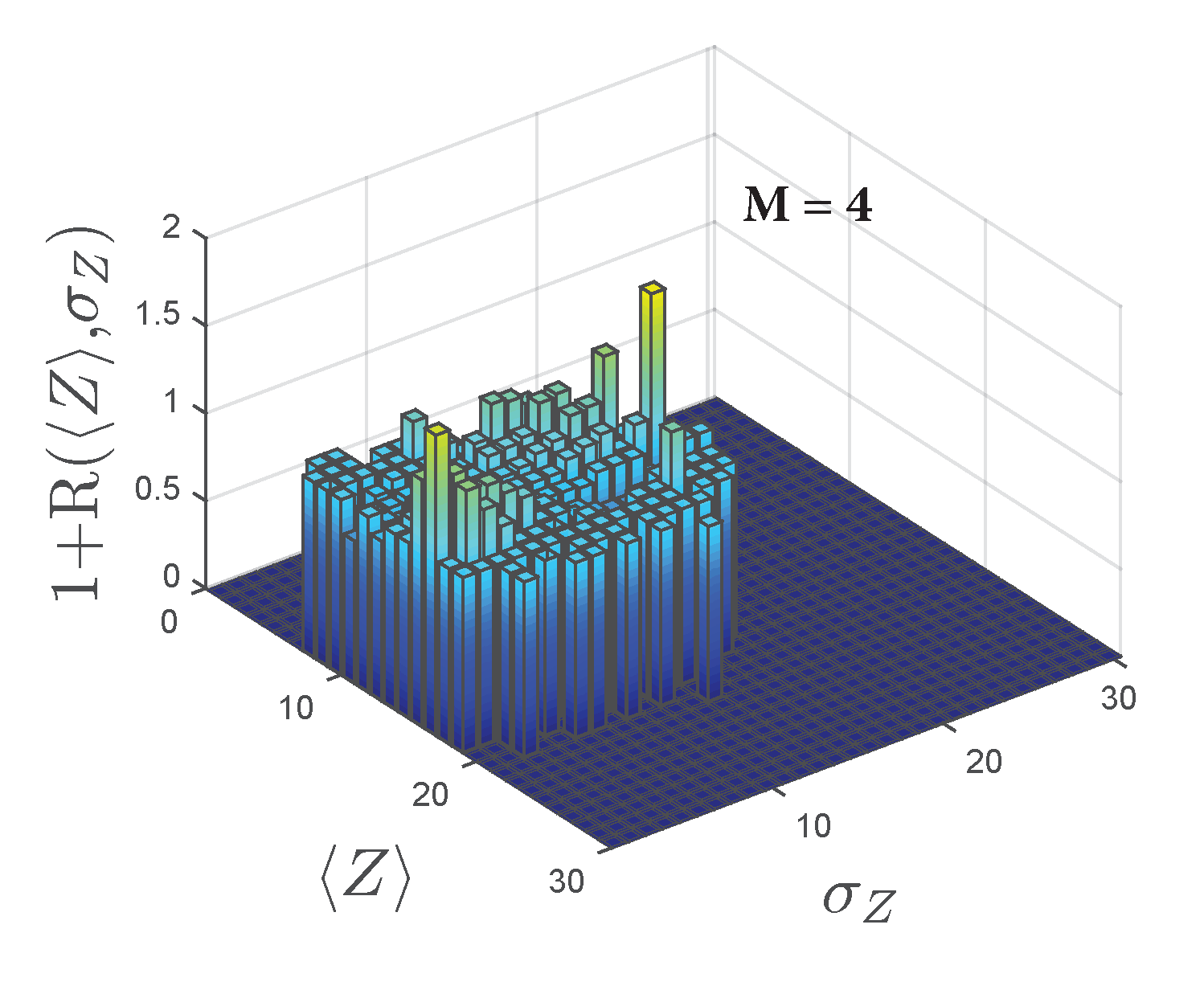}
\includegraphics*[width=0.48\textwidth]
{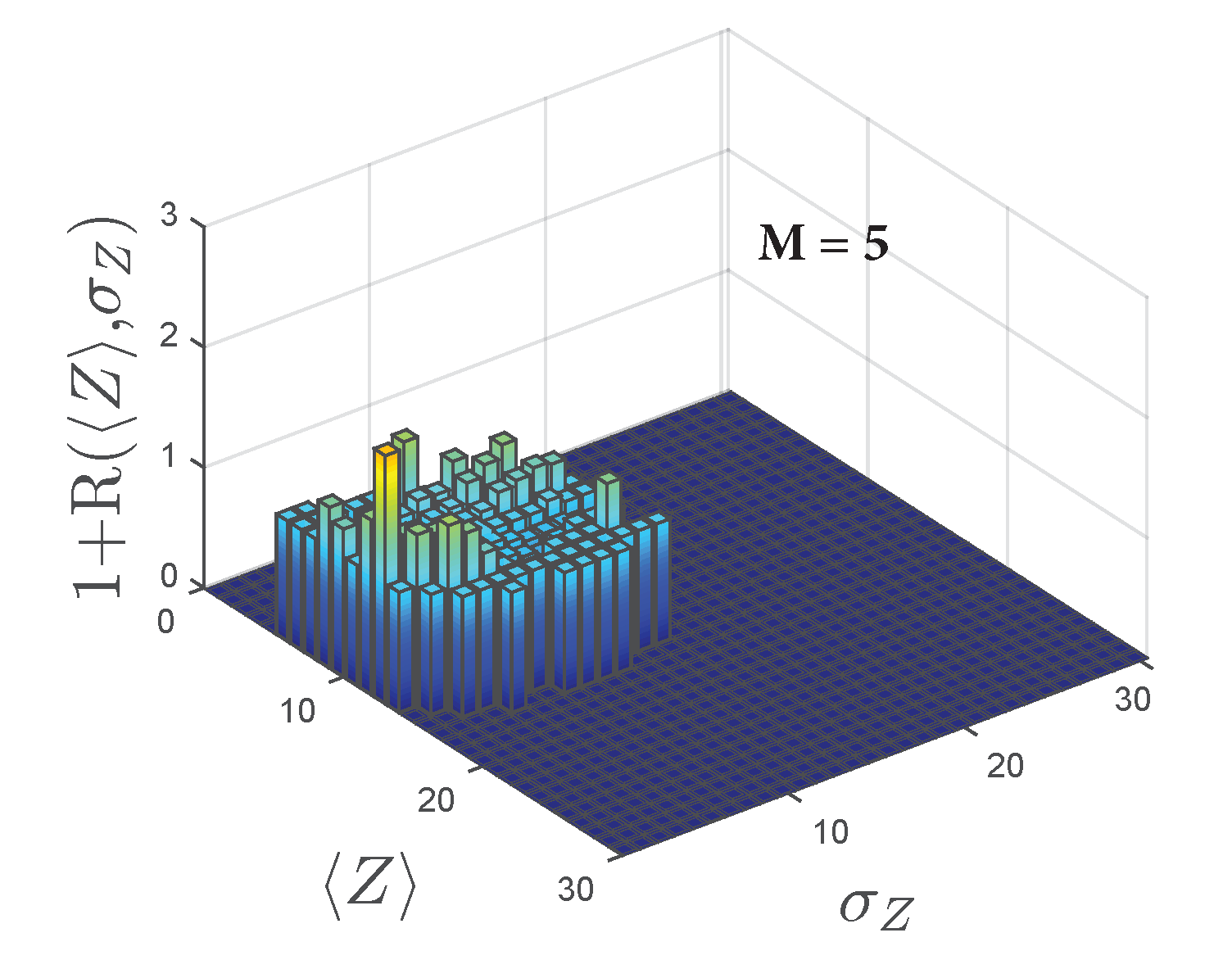}
\includegraphics*[width=0.48\textwidth]
{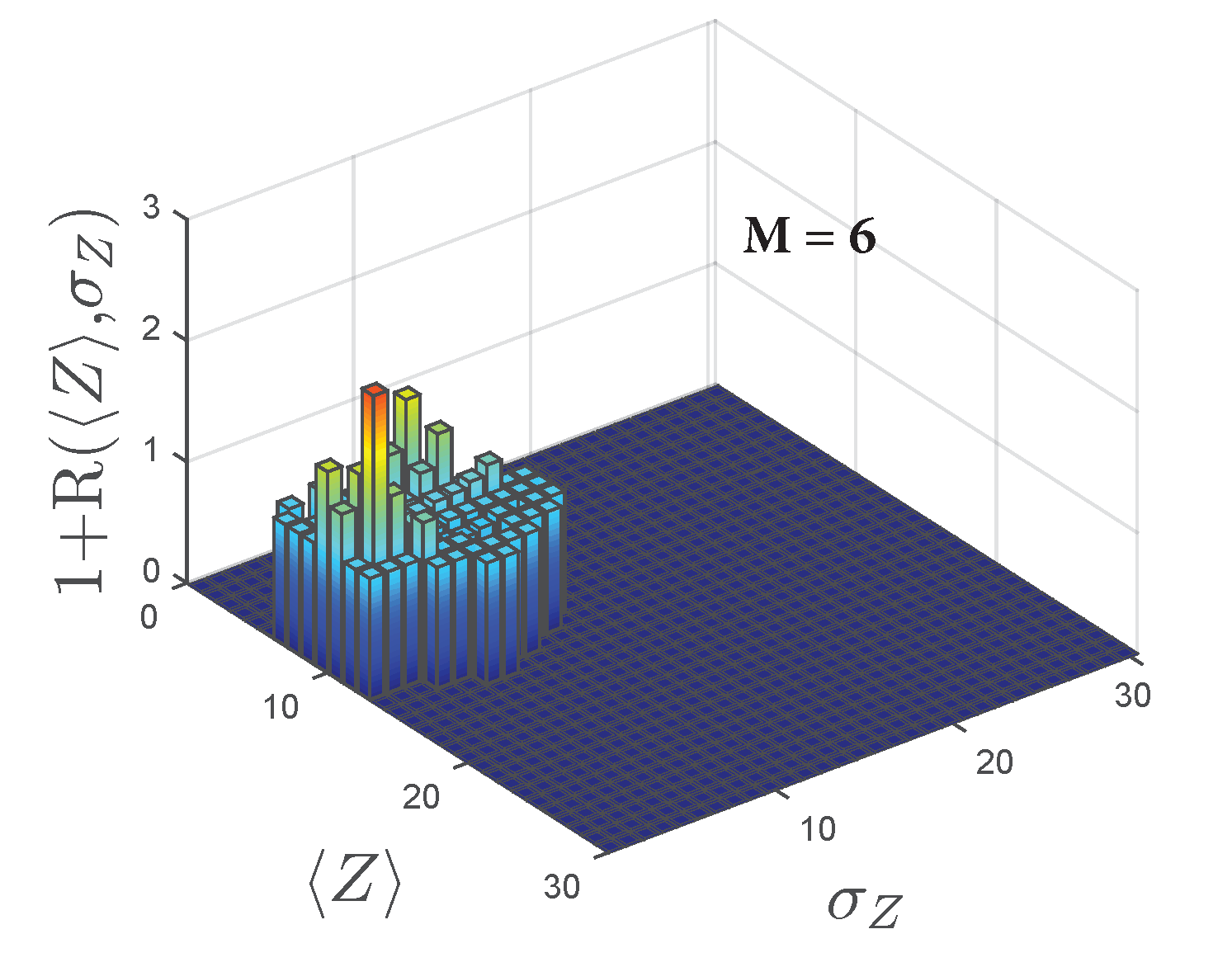}
\end{center}
\caption{(Color online) Experimental correlation functions for selected
quasifusion events
formed in central 32~\AM{} \nuc{124}{Xe}+\nuc{112}{Sn} collisions, for
fragment multiplicities 3 to 6. The color/gray scale is the same for all figures
and the maximum value of 3 for correlation functions corresponds to dark red/dark gray.}
\label{fig2}
\end{figure}

We shall present now the CF results for the
considered samples of events i.e. those with 3 to 6 fragments (Z$\geqq$5)
which correspond to higher statistics.
First of all, we must define what is the optimum variance bin $\sigma_Z$ 
bearing in mind that emitted primary fragments in these reactions 
have excitation energies of around 3 MeV per
nucleon~\cite{I39-Hud03,I66-Pia08}. 
Fig.~\ref{fig2} illustrates the CF measured for the different fragment
multiplicities with bins equal to one in $\Zmoy$ and $\sigma_{Z}$.
It concerns single source events selected at 32~\AM{} for the
\nuc{124}{Xe}+\nuc{112}{Sn} reaction. For clarity of presentation,
when the statistical error of the numerator, Y($\sigma_Z$,$\Zmoy$),
was greater than or equal to 50\%,  CF value was fixed to one. 
We observe significant peaks (CF values in the range 1.4 - 2.1) for
$\sigma_Z <$1 for each fragment multiplicity. For the
first time, the limited ranges of $\Zmoy$ contributing
to CF peaks are clearly observed (see also Table~\ref{table1}), which
verify what is expected for finite systems i.e. $M \times \Zmoy$
$\sim$ constant.
\begin{table}
\caption{Characteristics of events with $\sigma_{Z} < 2$ for the different
incident energies and reactions. For each fragment multiplicity M, the range of $\Zmoy$ 
contributing to the correlation peaks are indicated; $\Zmoy$ within
parentheses have CF values lower or equal to 1.1.}
\begin{center}
\begin{tabular}{lccccc}
\hline
   $M$ & & 3 & 4 & 5 & 6 \\
     E (AMeV) &reaction & & & &  \\
     \hline
32  &\nuc{124}{Xe}+\nuc{112}{Sn} &20 - 22 & 15 - 18 & 12 - 14 & 10 - 12 \\
32  &\nuc{136}{Xe}+\nuc{124}{Sn} &20 - 22 & 15 - 18 & 12 - 15 & 10 - 12 \\
45  &\nuc{124}{Xe}+\nuc{112}{Sn} &18 - 20 & 14 - 16 & 10 - 12 & 8 - 10 \\
45  &\nuc{136}{Xe}+\nuc{124}{Sn} &(16 - 20) & (12 - 16) & (10 - 14) &
(8-10) - 12 \\
\hline
\end{tabular} \\
\end{center}
\label{table1}
\end{table}
\begin{figure}
\begin{center}
\includegraphics*[width=0.96\textwidth]
{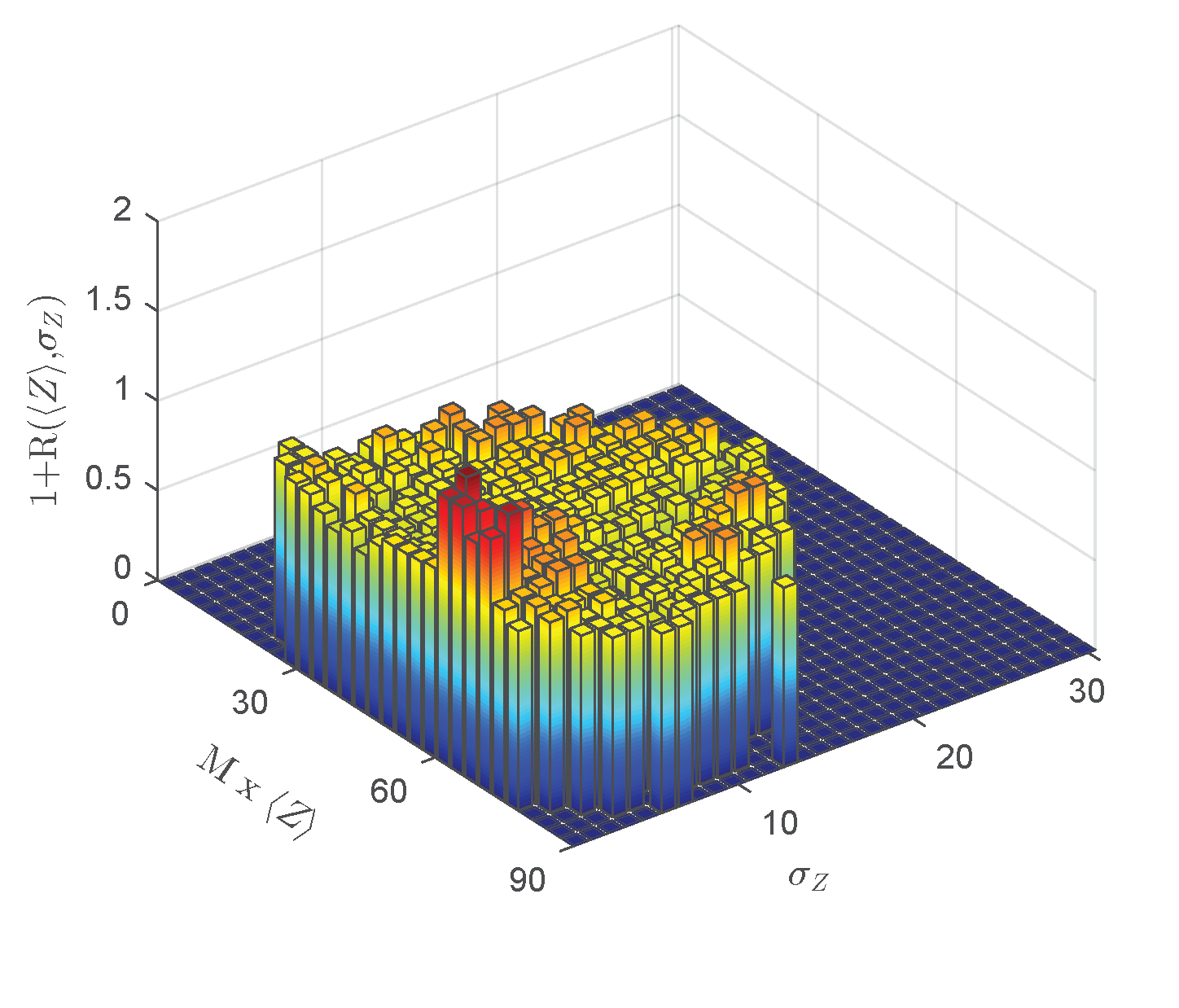}
\end{center}
\caption{(Color online) Experimental correlation function for selected
quasifusion events
formed in central 32~\AM{} \nuc{124}{Xe}+\nuc{112}{Sn} collisions.
Events with fragment multiplicities 3 to 6 are mixed. 
Correlation function is calculated for a $\sigma_{Z}$ bin equal to 1
 and a $ M \times \Zmoy$ bin equal to 3. The color/gray
scale for CFs has a maximum value for 1.6 which corresponds to dark red/dark gray.}
\label{fig3}
\end{figure}
\begin{figure}
\begin{center}
\includegraphics*[width=0.48\textwidth]
{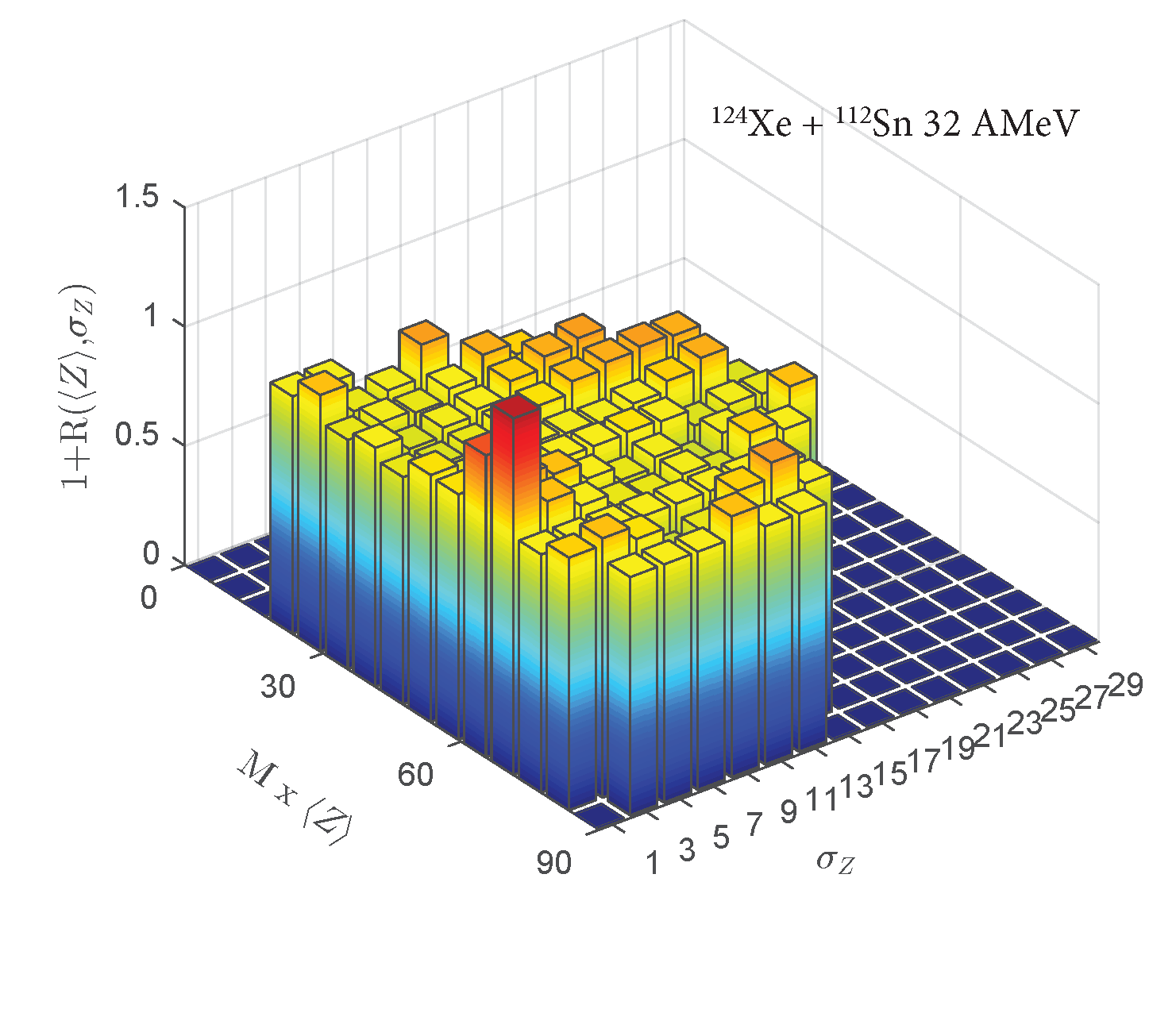}
\includegraphics*[width=0.48\textwidth]
{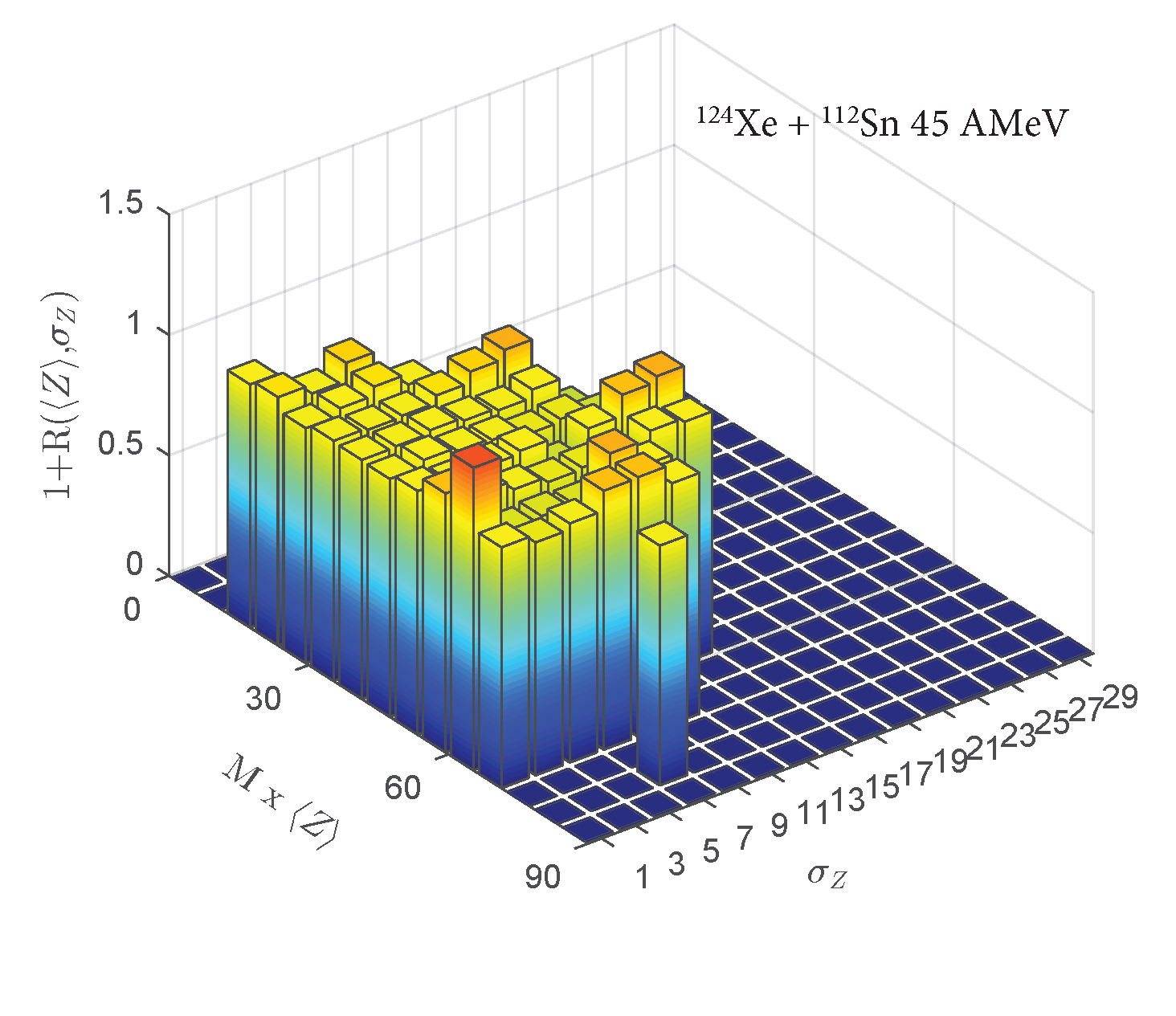}
\includegraphics*[width=0.48\textwidth]
{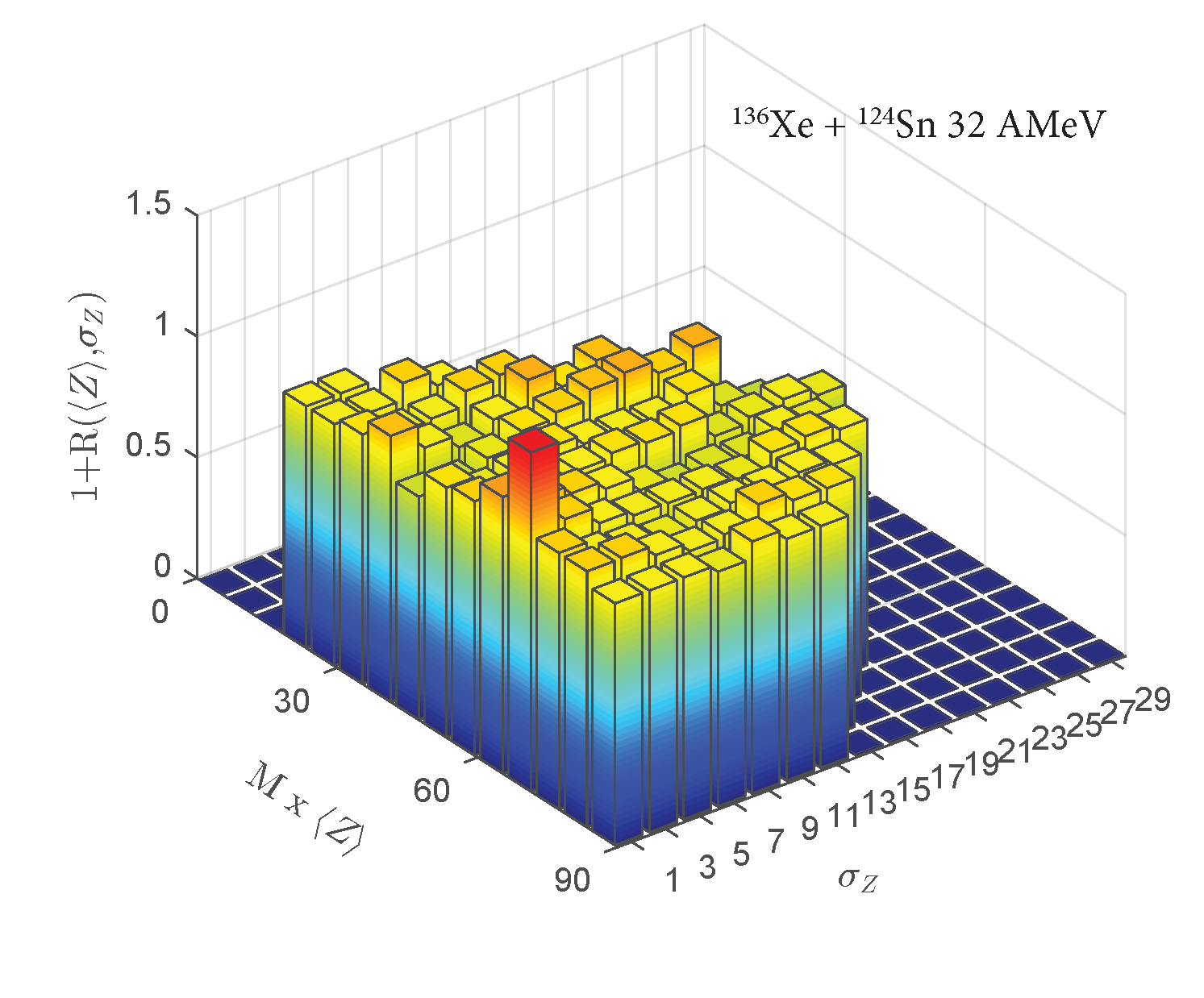}
\includegraphics*[width=0.48\textwidth]
{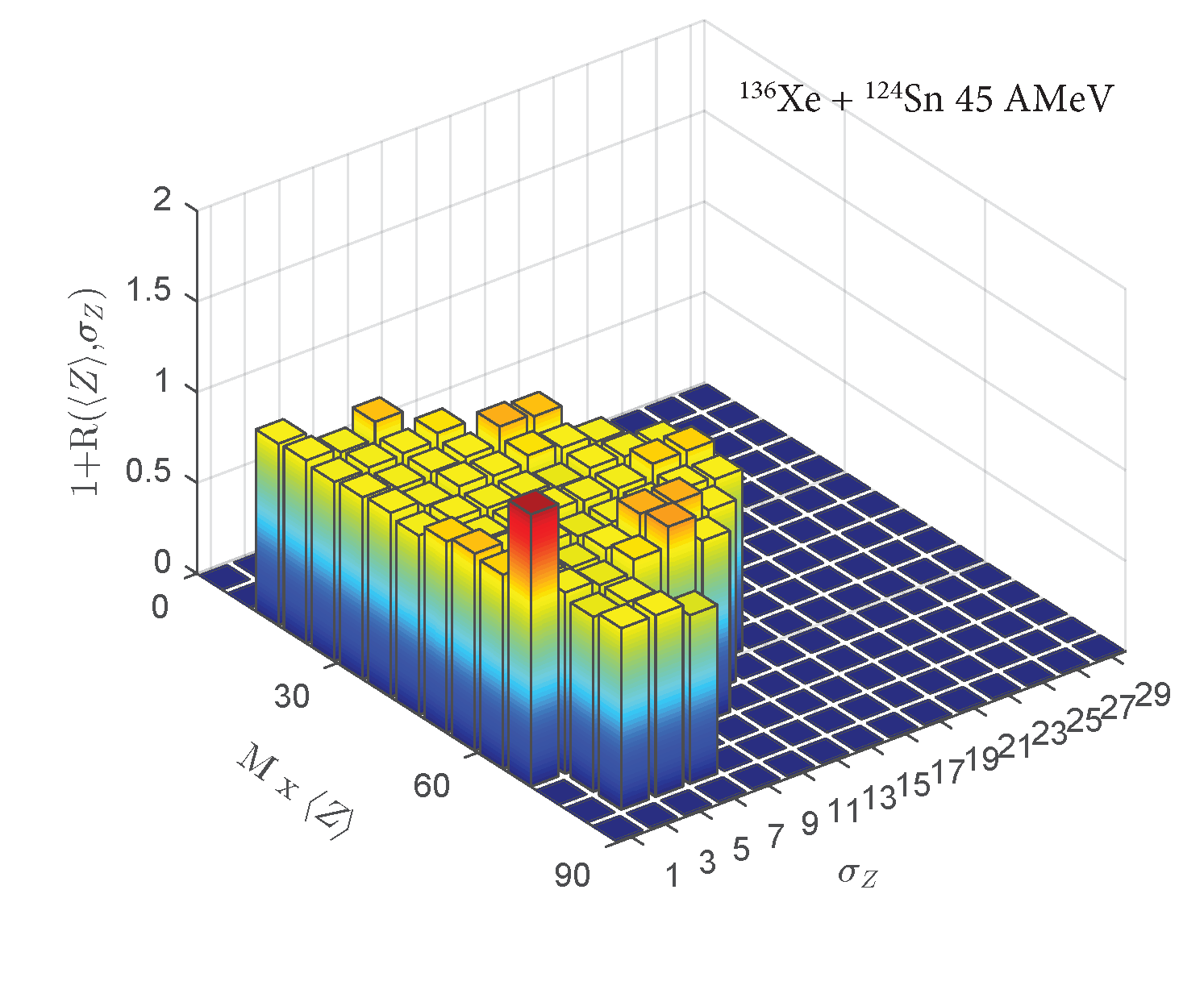}
\end{center}
\caption{(Color online) Experimental correlation functions for selected
quasifusion events
formed in central \nuc{124,136}{Xe}+\nuc{112,124}{Sn} collisions.
Events with fragment multiplicities 3 to 6 are mixed. 
Correlation functions are calculated for a $\sigma_{Z}$ bin equal to 2
 and a $ M \times \Zmoy$ bin equal to 6. The color/gray
scale for CFs has a maximum value for 1.6 which corresponds to dark red/dark gray.}
\label{fig4}
\end{figure}
We can then build CFs for all events,
whatever their multiplicity, by summing the correlated yields of all $M$ and
by replacing the variable $\Zmoy$ by  $ M \times \Zmoy$ .
Uncorrelated yields are  constructed and weighted in proportion to real
events of each multiplicity.
In Fig.~\ref{fig3} the resulting CF is displayed; the $ M \times \Zmoy$
bin is now equal to three. As for Fig.~\ref{fig2}
CFs have been fixed to one when the statistical error on the
correlated yield was equal or greater than 50\%. As expected a limited region 
with CF values greater than one (1.34 - 1.49) is
observed at very low $\sigma_{Z}$ ($<$ 1) for $ M \times \Zmoy$ bins
60-63, 63-66, 66-69 and 69-72.
We also highlight with the help of the present figure a complementary
contribution coming from $\sigma_{Z}$ (1 - 2) with CF values (1.31 -
1.57) equivalent to the previous ones that one must take into account. This fixes the
upper limit at 2  for the $\sigma_{Z}$ of events with nearly
equal-sized fragments.
The complementary contribution comes from the broadening of the
fragment $Z$ distribution introduced by the deexcitation of primitive
fragments. On the observed $Z$ range for fragments,
simulations of their deexcitation
from 3 MeV per nucleon excitation energy~\cite{I39-Hud03,I66-Pia08}
generate an extra $\sigma_{Z}$ value of around one unit~\cite{Dur92,Cha10}.
Note that
two other regions with CF values greater than one appear in the figure.
A first one, at low $ M \times \Zmoy$ values, 
which corresponds to quasifusion-evaporation events with
emission of several light fragments associated to the evaporation
residues and a second, at high $ M \times \Zmoy$
values, coming from quasifusion-fission events accompanied by one or
two light fragments. 
Fig.~\ref{fig4} (upper left) results from  what we
learned from Fig.~\ref{fig3} i.e. to use larger bins in both $\sigma_{Z}$
(2 units) and $ M \times \Zmoy$ (6 units). We observe only two CF peak at low $\sigma_{Z}$
with values 1.27 and 1.50 which cover the $ M \times \Zmoy$ range 60-72.
Their confidence levels in sigma units, obtained from the deviation to
1, $R(\sigma_Z, \Zmoy)$, of the correlation function, normalized to its
statistical error bar, $\sigma_{1 + R(\sigma_Z, \Zmoy)}$, calculated from
the numerator only are respectively equal to 6.09 and 7.29, which
definitively establishes
the presence of spinodal fluctuations.
\begin{table}
\caption{CF peaks with $\sigma_{Z} < 2$: values and confidence levels for the
different incident energies and reactions.}
\begin{center}
\begin{tabular}{lccccc}
    \hline
     E (AMeV) &reaction &$ M \times \Zmoy$ range&CF value &conf. level (sigma units) \\
     \hline
32  &\nuc{124}{Xe}+\nuc{112}{Sn} & 60-66 & 1.268 & 6.09\\
    &                            & 66-72 & 1.498 & 7.29\\
32  &\nuc{136}{Xe}+\nuc{124}{Sn} & 60-66 & 1.143 & 2.98\\
    &                            & 66-72 & 1.402 & 6.47\\				 
45  &\nuc{124}{Xe}+\nuc{112}{Sn} & 54-60 & 1.075 & 2.29\\
    &                            & 60-66 & 1.257 & 0.82\\
45  &\nuc{136}{Xe}+\nuc{124}{Sn} & 66-72 & 1.502 & 2.08\\
\hline
\end{tabular} \\
\end{center}
\label{table2}
\end{table}

The procedure described above was followed for the three other studied systems.
We see in Table~\ref{table1} that, whatever the incident energy and
the reaction, the same 
or very similar $\Zmoy$ ranges are contributing to the
correlation peaks for the two reactions. We also note that, except for
the highest multiplicity, low CF values
were measured for the neutron rich system at 45~\AM{} incident energy,
which makes more difficult a precise determination of the $\Zmoy$ range. 
Fig.~\ref{fig4} summarizes the results. 
As for Figs.~\ref{fig2} and~\ref{fig3},
CFs have been fixed to one when the statistical error on the
correlated yield was equal or greater than 50\%.
For the four systems we observe
CF peaks at low $\sigma_{Z}$ with values varying from 1.08 to 1.50 and
confidence levels from 0.82 to 7.29 sigma units (see also
Table~\ref{table2}).
Covered  $ M \times \Zmoy$ domains  are the same (60 - 72) for both
reactions at 32~\AM{} incident energy whereas at higher incident energy
the neutron rich system covers a range a little bit higher (66 - 72)
than the neutron poor one (54 - 66). Finally the percentages of events
($Y(\sigma_Z, \Zmoy)$/total number of events)
and extra events ($Y(\sigma_Z, \Zmoy)$ is replaced by 
$Y(\sigma_Z, \Zmoy)$ - $Y'(\sigma_Z, \Zmoy)$) measured  are reported in
Table~\ref{table3}.
\begin{table}
\caption{Numbers and percentages of events and extra events with $\sigma_{Z} < 2$
for the different incident energies and reactions.
Calculated errors are statistical.}
\begin{center}
\begin{tabular}{lccccc}
   \hline
     E (AMeV) &reaction &events &(\%) &extra events &(\%) \\
     \hline
32  &\nuc{124}{Xe}+\nuc{112}{Sn} & 1313 & 0.27 & 336 & 0.068 $\pm$ 0.004\\
32  &\nuc{136}{Xe}+\nuc{124}{Sn} & 1077 & 0.32 & 217 & 0.064 $\pm$ 0.004\\
45  &\nuc{124}{Xe}+\nuc{112}{Sn} & 1073 & 0.34 & 77 & 0.025 $\pm$ 0.003\\
45  &\nuc{136}{Xe}+\nuc{124}{Sn} & 68 & 0.030 & 15 & 0.0065 $\pm$ 0.0017 \\
\hline
\end{tabular} \\
\end{center}
\label{table3}
\end{table}
Within error bars, extra event percentages are
similar for both systems at the lower incident energy. At higher
incident energy we observe a strong reduction of percentages, in
particular for the neutron rich system. 
The observation of the reduction with incident energy,
at least for the more symmetric
system, is in good agreement with the negative heat capacity signatures
observed experimentally which fixed the upper limit of the coexistence zone (spinodal
region) in the
incident energy range 39-50~\AM{}~\cite{I46-Bor02}. This coherence
between both signals was not clearly observed from previous charge correlation
studies~\cite{I40-Tab03,Bor08}.
The large reduction of the signal observed for the neutron
rich system, a factor ten between 32 and 45~\AM{},  can be
understood in terms of the expected N/Z influence.
Indeed, if spinodal instabilities are at the origin of the dynamics
 of multifragmentation, as discussed before, theoretical
 calculations predict a reduction of 
 instabilities for asymmetric systems in relation with an increase of
 the instability growth time. More precisely, it is shown
 in~\cite{Col02} that, for $Sn$ isotopes, the most unstable modes
 associated to shorter instability growth times
 ($\simeq$50 fm/c) disappear when N/Z changes from 1.40 to 1.64. 
 If we consider that quasifusion systems produced by the collisions, with
 N/Z varying from 1.27 to 1.50, have to stay
 long enough in the spinodal region ($\sim$3 characteristic times) to allow
 important amplification of the initial fluctuations, one can
 qualitatively 
 understand the large extra reduction of the correlation signal for the
 neutron rich system at high incident energy as coming from
 insufficient reaction time. Such a situation also favors coalescence
 of primary fragments. Finally, note that the set of reaction trajectories
 in the density - temperature plane, close to the border of the
 spinodal region at 45~\AM{} can be slightly different for the two
 reactions.
 
In conclusion, two reactions \nuc{124}{Xe}+\nuc{112}{Sn} and
\nuc{136}{Xe}+\nuc{124}{Sn} have been used at two bombarding energies,
32 and 45~\AM{}, to produce quasifusion hot nuclei which undergo
multifragmentation. Using charge correlations the fossil signature of
spinodal instabilities i.e. the abnormal presence of equal-sized
fragments was definitively established at a confidence level of around
6 - 7 sigma units for both reactions at 32~\AM{} incident energy. 
At higher bombarding energy we observe a reduction of the signal
(by a factor around 3) for the more symmetric system. This reduction
is in good agreement
with the negative heat capacity signature previously observed which
fixed the upper limit of the spinodal region in the incident energy
range 39-50~\AM. The most important reduction of the signal is
registered for the neutron rich system (a factor around 10 between 32
and 45~\AM{} incident energies). This observation is consistent with the
theoretical expectation of an increase of the instability growth time
of spinodal fluctuations with an increase of N/Z for the hot nuclei.

The nature of the dynamics of the transition, i.e. the fragment
formation was the last missing piece of the puzzle concerning the
first order phase transition for hot nuclei. At this point one can
emphasize that multifragmentation
was early theoretically associated to a phase transition for nuclear
matter. Then multifragmentation became a generic term to name a
reaction where several fragments were observed in the exit channel.
From the present knowledge one can infer that the generic term covers
at least the reaction mechanisms involved in the phase transition
region and in the vaporization regime. We recall that one can locate
the onset of a pure gas phase at a thermal excitation energy of around
9-10 MeV per nucleon~\cite{Poc95,I72-Bon09,I79-Bor13}. It should be
also pointed out that, beyond the fragment production which we can
connect to spinodal instabilities and vaporization, the fragmentation
process may also arise from other dynamics like cavitation and nucleation.

Acknowledgements

One of the authors B.B. want to
underline the major role played by theoreticians of this domain
during the last twenty years of the previous century to give some guidance
to experimentalists equipped with powerful multidetectors.


\end{document}